\documentclass[twocolumn,english,prl,amssymb,aps,superscriptaddress,showpacs,twocolumn,amsmath,showkeys,floatfix]{revtex4-1}
\usepackage[T1]{fontenc}
\usepackage[latin9]{inputenc}
\usepackage{geometry}
\usepackage[active]{srcltx}
\usepackage{textcomp}
\usepackage{amsmath}
\usepackage{graphicx}
\usepackage{color}
\usepackage{esint}
\usepackage{svg}
\makeatletter
\usepackage{babel}

\makeatother

\begin{document}
	
	%\title{Nanolaser Mode-Locking of Hermite-Gaussian Modes}
	\title {Imaging  through fog using quadrature lock-in discrimination}
	\author{Shashank Kumar}
	\affiliation{Raman Research Institute, Sadashiv Nagar, Bangalore, India 560080}
	\author{Bapan Debnath}
	\affiliation{Raman Research Institute, Sadashiv Nagar, Bangalore, India 560080}
	\author{Meena M. S.}
	\affiliation{Raman Research Institute, Sadashiv Nagar, Bangalore, India 560080}
	\author{Julien Fade}
	\affiliation {Univ. Rennes, CNRS, Institut FOTON - UMR 6082, F-35000 Rennes, France}
	\author{Sankar Dhar}
	\affiliation {Shiv Nadar University, Gautam Buddha Nagar, Uttar Pradesh 201314, India}
	\author{Mehdi Alouini}
	\affiliation {Univ. Rennes, CNRS, Institut FOTON - UMR 6082, F-35000 Rennes, France}
	\author{Fabien Bretenaker}
	\affiliation{Raman Research Institute, Sadashiv Nagar, Bangalore, India 560080}
	\affiliation{Universit\'e Paris-Saclay, CNRS, ENS Paris-Saclay, CentraleSup\'elec, LuMIn, 91190 Gif-sur-Yvette, France}
	\author{Hema Ramachandran}
	\affiliation{Raman Research Institute, Sadashiv Nagar, Bangalore, India 560080}

	\begin{abstract} 
We report experiments conducted in the field in the presence of fog, that were aimed at imaging under poor visibility. By means of intensity modulation at  the source and two-dimensional quadrature lock-in detection by software at the receiver, a significant  enhancement of  the contrast-to-noise ratio was achieved in the imaging of beacons over hectometric distances. Further by illuminating the field of view with a modulated source, the technique helped reveal objects  that were earlier obscured due to multiple scattering of light. This method, thus, holds promise of aiding in various forms of navigation under poor visibility due to fog.  	\end{abstract}
	
	%\pacs{??}
	
	\maketitle
	
\section{Introduction and context}
Visibility is reduced in fog because the tiny droplets of water suspended in air cause random multiple scattering of light, thereby degrading the image-bearing capabilities of photons. This is detrimental to many imaging applications of optics in open air, based either on passive imaging of scenes immersed in fog, or on active detection of beacons. This latter situation is  of particular relevance when series of beacons are installed along runways to guide aircraft for landing and takeoff. It becomes impossible for the pilot to observe these beacons during thick fogs, and there is no other alternative in the case of airfields or aircraft not equipped with radio-frequency instrument landing systems. Similar problems exist in maritime navigation, railway transportation, and even for motor transport on highways. Such examples illustrate the need for a simple, cheap, and compact technique aimed at improving the visibility of optical beacons in foggy weather conditions, and if possible also viewing objects that do not themselves emit light. 

One class of ``fog-removal''  technique is purely computational, where image processing algorithms are used on single or multiple images to remove the effect of fog (e.g., \cite{Anwar2017} and references therein).
The other class of techniques exploits the physics of the problem and discriminates between different types of photon trajectories. Photons transiting a scattering medium are usually classified as (a) ballistic photons, that are forward scattered and retain their original direction of propagation, (b) snake photons, that are near forward scattered, and have paths that are not far from the ballistic,  and  (c) diffusive photons that are scattered through random angles over all directions and whose paths are scrambled such that the original direction of propagation is lost. The 
various approaches that have been used either select the small amount of ballistic (and snake) photons from among the huge amount of multiply scattered light that reaches the detector or the camera, or  exploit the diffusive light itself to gain imaging capabilities (see, for example, reviews \cite{Ramachandran1999,Dunsby2003}). As illustrations of the first strategy, the ballistic photons may be temporally discriminated using a pulsed source of light in conjunction with time-gated detection \cite{David2006} or time gated holography \cite{Kanaev2018}. An alternative technique also based on pulsed laser illumination exploits the different statistics of backscattered and reflected photons  \cite{Satat2018}. Many recent works have aimed at  imaging or focussing light through strongly scattering media, mainly for bio-medical imaging through live tissues \cite{Popoff2010, Katz2012, Bertolotti2012}. These techniques, that characterise the scattering properties of the turbid medium, have limited applicability to fog, as the scatterers in fog are constantly moving at high speed.

In the context of imaging through fog, another simpler and cheaper approach consists in using a modulated continuous-wave source of light and relies on the fact that the intensity variation of the  ballistic photons retains a phase relationship with the intensity modulation of the  source, while that of the  multiply scattered photons does not. 
This technique requires a demodulation of the detected signal at the modulation frequency.  High modulation frequencies are required for true ballistic filtering \cite{Panigrahi2016}, especially when the scatterers have a large anisotropy factor with significant forward scattering, as is the case with Mie scatterers. Rayleigh scatterers, due to their almost isotropic scattering, should in principle permit discrimination of the diffusive  from the ballistic low scattering order components at lower modulation frequencies \cite{Panigrahi2016}, however at the cost of a reduced number of ballistic or snake-like photons. Thus, the modulation-demodulation technique has proven to be efficient in enhancing source visibility, by discriminating against the background contributed by ambient lighting, sources modulated at different frequencies, and to varying extents, the diffusive photons. The demodulation may be performed using a bucket detector followed by lock-in electronics. This requires a step scan of the detector to build a two-dimensional image \cite{Fabien}, and is thus time-consuming. Demodulation may be  performed numerically  using Fourier transform over a time-series of images \cite{HemaOC}.  Though many optimised algorithms are available for fast Fourier transform, the technique  has its drawbacks \cite{Sudarsanam2016} - it requires large memories to store the time series (1K frames or more) of images (each megapixels or more), on-camera buffer sizes are not large enough, and  storage and read-out of images on the computer takes time. An alternative technique that was recently demonstrated  consists in performing this demodulation instantaneously by optical means, with promising perspectives of high-frequency operation \cite{Panigrahi2020}. This, however, comes with increased complexity and cost of the optical elements. A different and simpler approach, suited for moderate frequencies, has also been recently demonstrated \cite{Sudarsanam2016}. It consists in performing quadrature lock-in discrimination (QLD) \cite{Mullen2007} computationally to obtain real-time demodulation of images. By multiply-and-accumulate operation on each image as it is acquired, the need for storing a series of images is eliminated. By multiplying by the two quadratures of the modulation, the need for phase matching  between the source and receiver is  obviated. Exploiting the  task and data parallelization capabilities of present-day desktop computers, this technique has been shown to lead to real-time display of $600\times 600$ pixel images with low latency and at rates faster than the eye bandwidth. However, till now, this technique has been applied only in table-top experiments where suspensions of microspheres were used as the scattering medium. The aim of the present paper is to test this technique in actual field conditions in real fog.

\section{Principle of QLD}
\begin{figure}[htbp]
\centering
\includegraphics[width=1.0\linewidth]{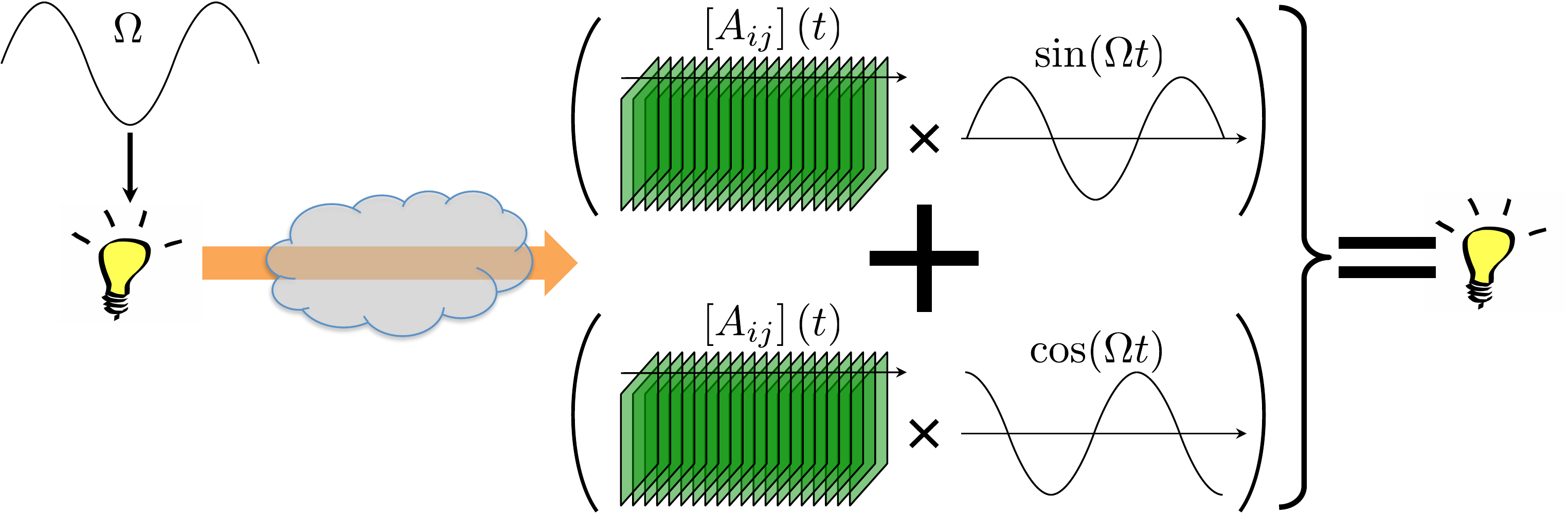}
\caption{Principle of quadrature lock-in discrimination.}
\label{Fig01}
\end{figure}
The principle of QLD is schematized in Fig.\,\ref{Fig01}. We start from a source whose intensity is modulated at angular frequency $\Omega$. After propagation in the scattering medium, a series of frames are acquired by the camera at a rate much larger than $\Omega/2\pi$ and transferred in real time to a computer. The computer multiplies two copies of each frame by two sinusoidal signals at frequency $\Omega$, which are in quadrature one to the other. After time averaging and addition, this allows the retrieval of the amplitude of the modulated part of the intensity that falls on each pixel and  eliminates the DC background that blurs the original raw frames. Thanks to the fact that the modulation at $\Omega$  is imprinted only on the light emitted by the modulated beacon, this technique permits the reduction of the noise in the acquired frames that comes from background light. However, the modulation frequency is too low to allow to filter out the part of the modulated light which is multiply scattered \cite{Mullen2009}. Thus, this technique does not perform a real selection of ballistic photons only.

\begin{figure}[htbp]
\centering
%\includegraphics[width=0.8\linewidth]{Fig01.pdf}
%\bigskip
\includegraphics[width=1.0\linewidth]{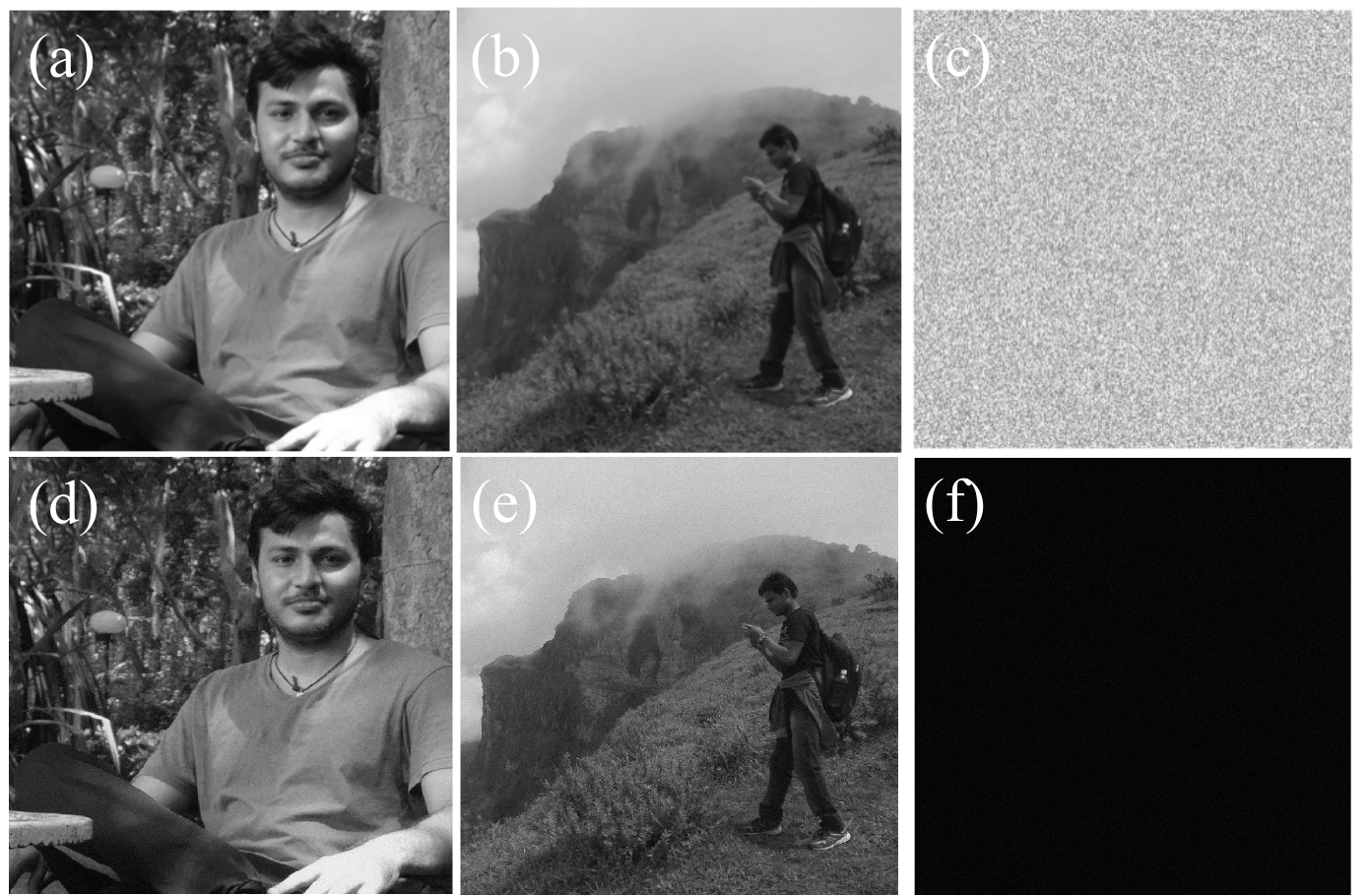}
\caption{(a) Image to be modulated at 13~Hz. (b) Image to be modulated at 17~Hz. (c) One of the frames of the mixed image after addition of noise. Original images are not visible at all. The retrieval of original image is done from 1105 such frames. (d) Image obtained upon demodulation at 13~Hz.  (e) Image obtained upon  demodulation at 17~Hz. (f) Image obtained upon demodulation at  14~Hz.}
\label{Fig02}
\end{figure}
We give a  simulation of this principle. We begin with two different 8-bit images (see Figs.\,\ref{Fig02}(a,b)). We sinusoidally modulate the first image at $\Omega/2\pi=13\,\mathrm{Hz}$ and the second  image at $\Omega/2\pi=17\,\mathrm{Hz}$. To obtain entire numbers of period, we need at least $13\times17=221$ frames. We then multiply this number of frames by 5 to increase the number of frames, leading to a total number of $5\times13\times17=1105$ frames for each image. An offset of 255 is added to every frame pixel in order to make all the values positive. The frames of the two series are then added. Finally, we mimic the noise by adding to every pixel of every frame a random number uniformly distributed between 0 and $5\times 255= 1275$. This represents the raw frames that a camera would record when simultaneously viewing the two differently intensity modulated sources through a randomly scattering medium.  One such simulated raw frame is shown in Fig.\,\ref{Fig02}(c). Clearly, neither of the original images of Figs.\,\ref{Fig02}(a) and \ref{Fig02}(b) is visible. We then apply QLD in order to retrieve the original images from the series of raw frames. The results of processing at 13~Hz, 17~Hz and 14~Hz are shown in Figs.\,\ref{Fig02}(d-f), respectively. The original images are retrieved, with the added noise being filtered out when processed at the "correct" frequency. On the other hand, QLD performed at a ``wrong'' frequency reveals neither image.

\begin{figure}[htbp]
\centering
\includegraphics[width=1\linewidth]{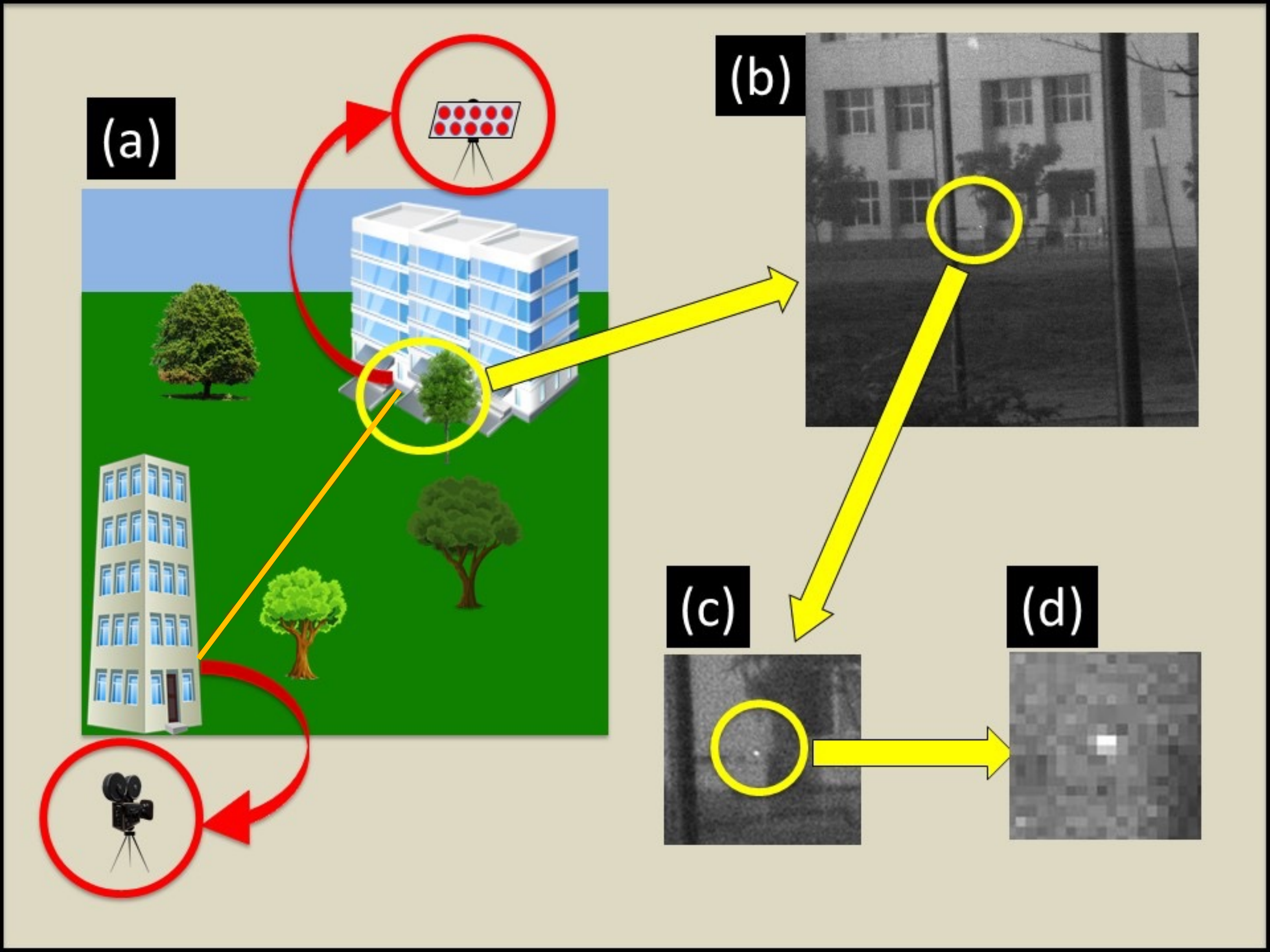}
\caption{{\color{black}(a) Schematic of the experiment performed.The red curved arrows indicate the location of the camera in the  building in the foreground, and the LED panel fixed on the building at the background. The distance between the two is 150~m; the orange full line indicates the clear line-of-sight between them. (b) An image of a portion of the building housing the LED panel, as captured at daybreak, in the absence of fog. The portion circled in yellow is enlarged and shown in (c); this is further enlarged and shown in (d). The processed images appearing in later figures may be compared with (d).}}
\label{Fig03}
\end{figure}

\section{Field experiments}
\subsection{Application of QLD to a modulated beacon}
To test the efficiency of  QLD in real fog, we performed field experiments as schematized in Fig.\,\ref{Fig03}, over a period of two months during peak winter at Shiv Nadar University, Uttar Pradesh.  A LED panel, consisting of 10 uncollimated LEDs connected in parallel on a 10\,cm x 16\,cm standard printed circuit board, emitting typically 1~Watt each in the red (around 640~nm), was used as the source of light.  Several factors, apart from the ease of availability of LEDs,  contributed to this choice of wavelength. Red light is conventionally used to signify danger, and most warning lights are in this colour. Further, the Rayleigh scattering of light is proportional to the inverse fourth power of wavelength, and therefore the red part of the spectrum should be preferred over the blue. Finally, most cameras have the highest sensitivity in this part of the spectrum.  The entire bunch of LEDs were so closely spaced that they could not be individually resolved at the camera, and thus appeared as a single bright source.  The current through the LEDs was  modulated (peak-to-peak modulation amplitude equal to 30\,\% of the average current) so that the intensity of the emitted light could be varied  sinusoidally at any  frequency in the range 13 - 17 Hz. The detector used was a 16-bit Andor Neo sCMOS camera, controlled by Andor Solis software, with a 8-48\,mm F/1-1.2 zoom lens from Ernitec. In the conditions of our acquisitions, the actual pixel dynamic of the camera is 13.4~bits, obtained from the ratio of the pixel well depth (30000 electrons) to the RMS read noise (2.8 electron according to the manufacturer). Series of frames of the scene at the desired frame rates were acquired  and transferred to a desktop computer where they were  processed for the  extraction of images using the QLD technique. The distance between the source and the detector was 150 meters. We used natural features of the scene to evaluate visibilities, which  during our acquisitions, ranged  between 30 to 150~m.

\begin{figure*}[htbp]
\centering
\includegraphics[width=1.5\columnwidth]{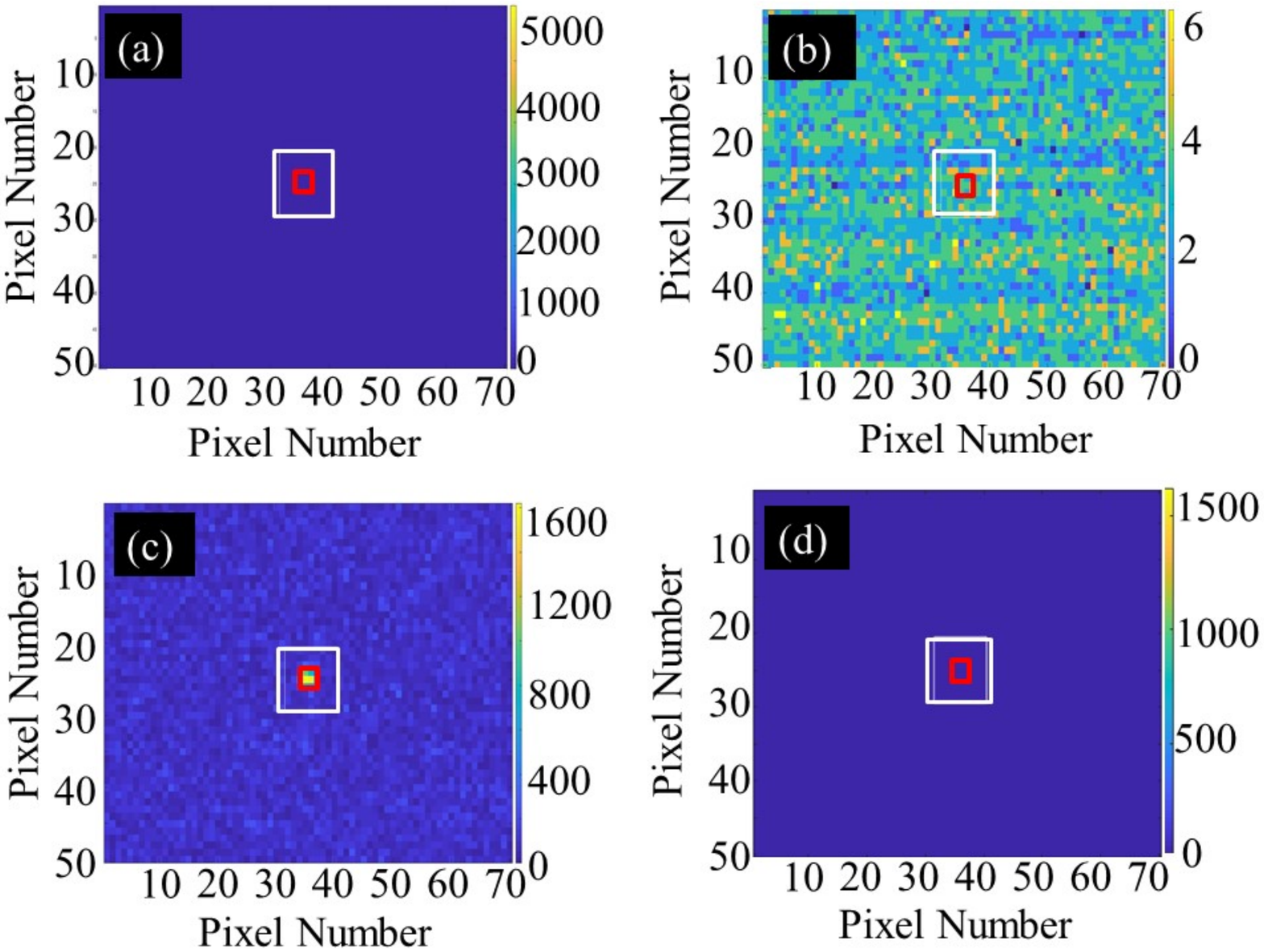}
\caption{Examples of raw and QLD-processed images acquired at day-break with a visibility of 40~m. (a) One full-scale raw frame ($\mathrm{CNR}=2.3$, 1~ms exposure time). (b) Same as (a) for a reduced color scale. (c) Corresponding processed image obtained from 10140 raw frames acquired at 390 frames per second. CNR is now equal to 11.0 (d) The image obtained by averaging all 10140 raw frames. The source is not seen, the CNR is 2.5. {\color{black} The red and the white squares in the figures represent the "object" and the "background" regions defined after Eq.\,(\ref{eqCNR}).} }
\label{Fig04}
\end{figure*}

In Fig.\,\ref{Fig04} we describe a set of recordings made when the source was modulated at 13~Hz. Each data set  was based on the acquisition of a total of 10,140 frames collected at rates of 260 or 390~Hz, and with exposure times per frame ranging from  0.5~ms to 5~ms depending on the weather condition. A raw frame recorded in one of the experiments is shown in false color in  Fig.\,\ref{Fig04}(a). Although this frame was acquired at day-break, the LED panel, located at a distance of 150~m, is not visible because of the heavy fog conditions with visibility of 40~m. This frame is re-plotted in Fig.\,\ref{Fig04}(b) with a different color scale corresponding to a scale enhancement by a factor 850. The  LED panel is still almost impossible to distinguish. After QLD processing of 10,140 such raw frames, we obtain the result reproduced in Fig.\,\ref{Fig04}(c), where, in contrast to Figs.\,\ref{Fig04}(a,b), one can now clearly see the LED panel at the centre.  Figure\,\ref{Fig04}(d) shows the image obtained upon averaging the 10140 raw frames, without QLD processing. The source is not visible in this case. 

In order to gain a more quantitative picture of the improvement of the image quality obtained using QLD, we measure the Contrast-to-Noise Ratio (CNR), defined as :
\begin{equation}
\mathrm{CNR}=\frac{\langle I_{\mathrm{obj}}\rangle-\langle I_{\mathrm{back}}\rangle}{\left\langle\left(I_{jk}-\langle I_{\mathrm{back}}\rangle\right)^2\right\rangle_{\mathrm{back}}^{1/2}}\ .\label{eqCNR}
\end{equation}
In this expression, $\langle I_{\mathrm{obj}}\rangle$ is the average value of the signal over the pixels covering the modulated source. In the case of Fig.\,\ref{Fig04}, it corresponds to the $3\times3$ pixels surrounded by the red rectangle. The quantity $\langle I_{\mathrm{back}}\rangle$ is the average of the signal recorded in the background surrounding the object. In Fig.\,\ref{Fig04}, it corresponds to the 8 blocks of $3\times3$ pixels contained in the white rectangle. In the denominator, the averaging is taken over the same $i,j$ pixels belonging to the surrounding background, so that this denominator is the square root of the variance of the background noise.

\begin{figure}[]
\centering
\includegraphics[width=1.0\columnwidth]{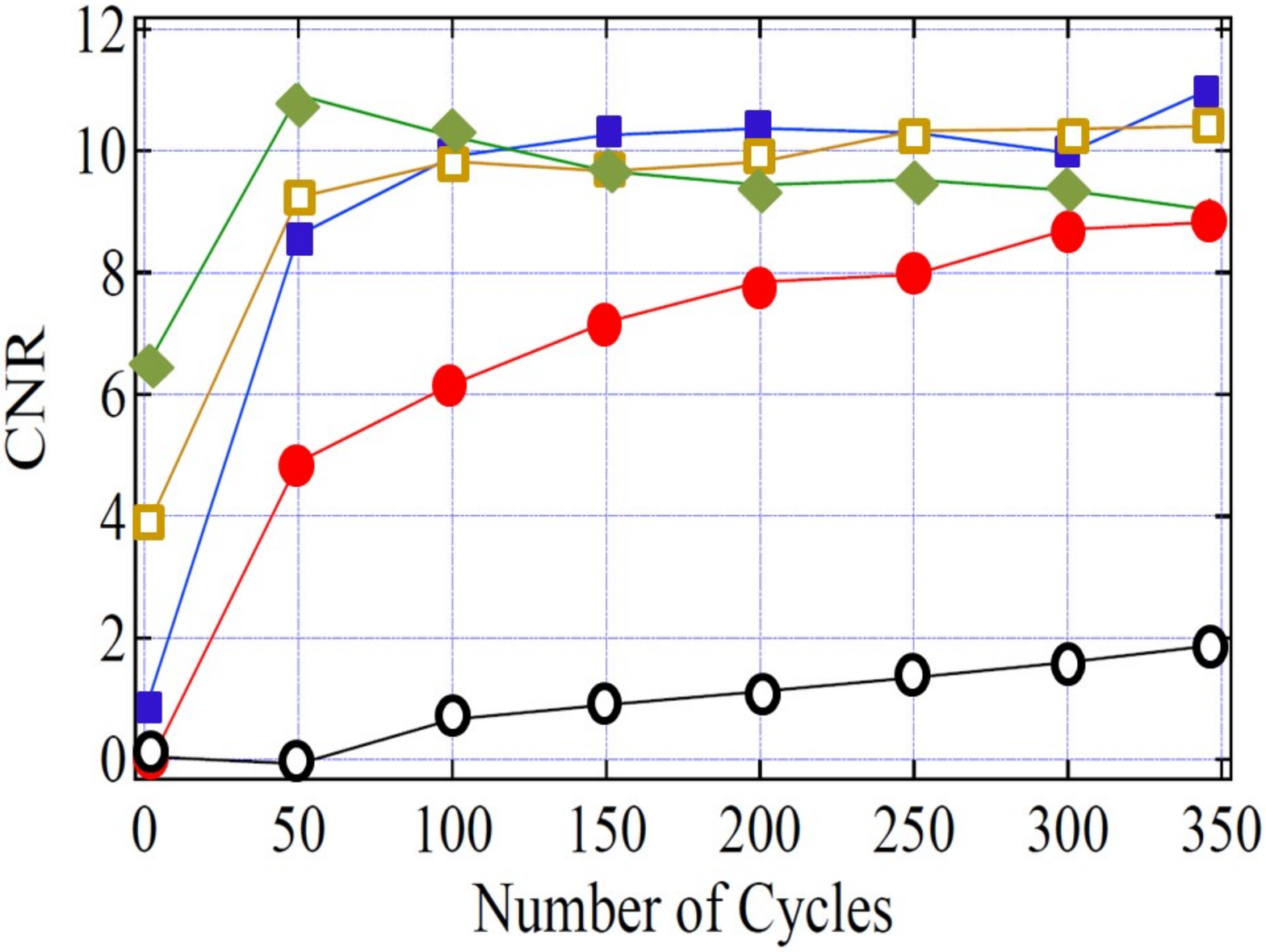}
\caption{Evolution of the CNR as a function of the number of processed modulation cycles. $\mathrm{Modulation\ frequency}=13\,\mathrm{Hz}$; 390 frames per second. Full circles: 40~m visibility; 1~ms exposure time. Full squares: 40~m visibility; 1~ms exposure time. Full diamonds: 50~m visibility; 2~ms exposure time. Open circles: 60~m visibility; 1~ms exposure time. Open squares: 60~m visibility; 0.5~ms exposure time.}
\label{Fig05}
\end{figure}
 For the data of Fig.\,\ref{Fig04} (visibility 40~m) QLD applied to images acquired at a distance of 150~m from the source permits to increase the CNR from 2.3 to 11. In Fig.\,\ref{Fig05}, we plot the evolution of the CNR as a function of the number of cycles over which QLD averaging is performed. The results in this figure, obtained in five sets of experiments performed at day-break for visibilities ranging from 40 to 60~m, 
 %Depending on the acquisitions, the CNR starts from values ranging from almost 9 to 7. 
show that the CNR can be significantly increased using the technique of QLD. Four of them permit to reach a value of the CNR larger than 8, typically with the number of periods needed to optimize the CNR being of the order of 100. While the general trend for all data sets is the same, variations exist. For example, the curves shown in full circles and in full squares,  both of which are for a visibility of 40~m, are  different. This may be attributed to the somewhat different nature of the fog on the two days the data was taken, or to possible variations of the ambient illumination. It is well known that fog can have droplets with sizes ranging from sub-micrometer to several micrometers \cite{NASA}, and thus the scattering can vary from being isotropic to significantly anisotropic (forward scattering), leading to a difference in the efficiency of the QLD technique.  This effect is more pronounced in the curve with open circles in Fig.\,\ref{Fig05}, which is quite different from the other four : the CNR is initially close to zero, and increases very slowly as a function of the number of modulation cycles over which QLD is performed, reaching a modest value of 1.9 even when all the available data are processed. Though this set of data was not obtained for a visibility significantly lower than  the other sets of data, the nature of fog varied  during observation. The wind was relatively strong then, and fog trails could be seen passing across the scene, indicating that the density and diameter of the water droplets was changing with time during the acquisition time, thus leading to the different results for seemingly identical conditions. It is also quite possible that these variations in the fog during acquisition explains why, in the case of the data represented as full diamonds, the CNR slowly decreases when the number of cycles is increased above 100.

\begin{figure*}[]
\centering
%\vspace{-0.3cm}
%\includegraphics[width=0.8\linewidth]{Fig06N1.pdf}
\includegraphics[width=1.5\columnwidth]{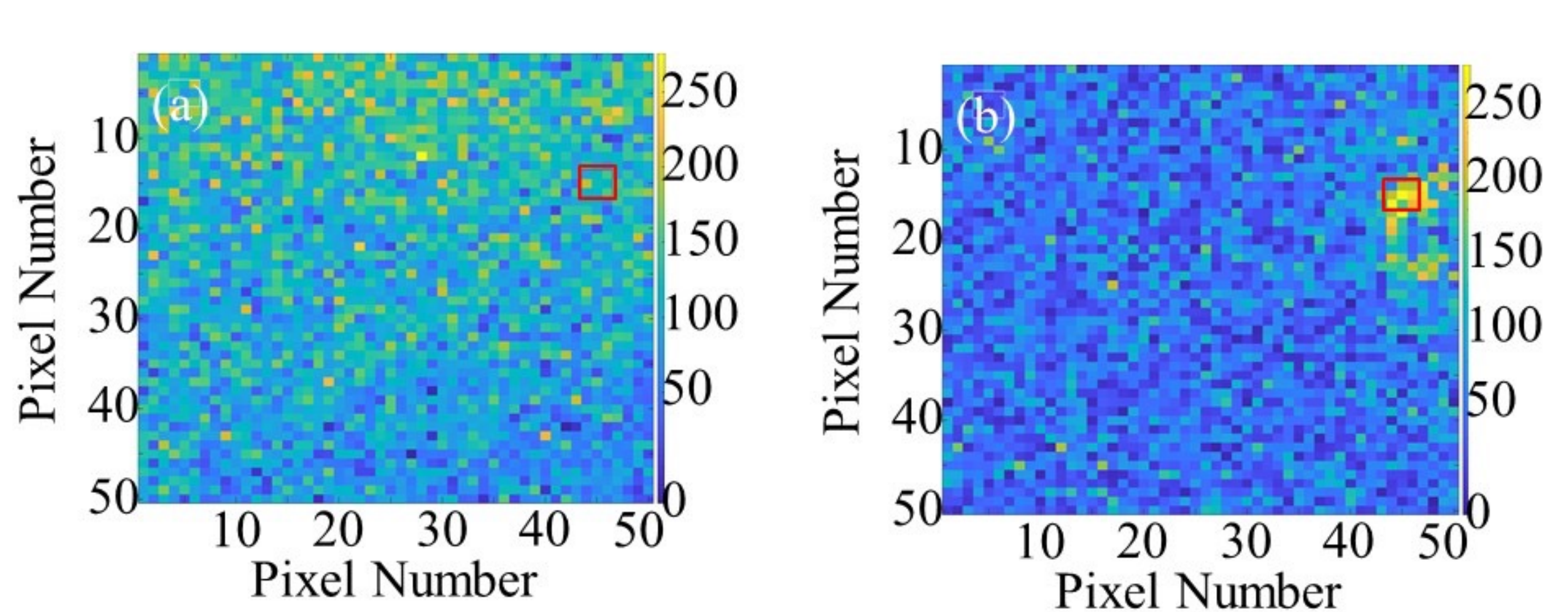}
%\vspace{-3.5cm}
\caption{Examples of raw and QLD-processed images acquired at day-break with a visibility of 30~m. The red rectangle shows the position of the piece of cardboard that is illuminated by the light of modulated LED panel. (a) One full-scale raw frame (5~ms exposure time). (b) Corresponding processed image obtained from 10,400 raw frames acquired at 160 frames per second.  The modulation frequency is 16~Hz.}
\label{Fig06}
\end{figure*}

\subsection{Application of QLD to an illuminated object}
 We have, so far, focused on imaging light beacons through fog using the technique of QLD.  We now investigate whether this method can be used under foggy conditions to enhance the visibility of an object that is illuminated by the modulated source of light, rather than the source of light itself. The data presented in Fig.\,\ref{Fig06} were obtained at day-break by illuminating using the modulated LED panel,  a piece of cardboard located on the right of the picture.  The distance between the LED panel and the illuminated cardboard was about 20\,cm. The distance between this object and the camera was  75~m while the visibility through fog was estimated at about 30~m. In the raw frame of Fig.\,\ref{Fig06}(a)  the object cannot be distinguished;  the associated CNR, equal to 0.3, is indeed quite low. However, after QLD processing of 10,400 such frames acquired at a rate of 160 images per second,  the image of Fig.\,\ref{Fig06}(b) was obtained, where the object illuminated by the LED panel can be clearly distinguished from the noisy background, with a  CNR of  1.9. Notice that in the present experiment the QLD technique works at low modulation frequencies because the modulated light source is close to the illuminated object. Making it to work with the modulated light source close to the camera and thus far from the object would probably require much higher frequencies or even pulsed illumination.

\subsection{QLD imaging in daylight condition}
\begin{figure*}[]
\centering
\vspace{-0.2cm}
\includegraphics[width=1.5\columnwidth]{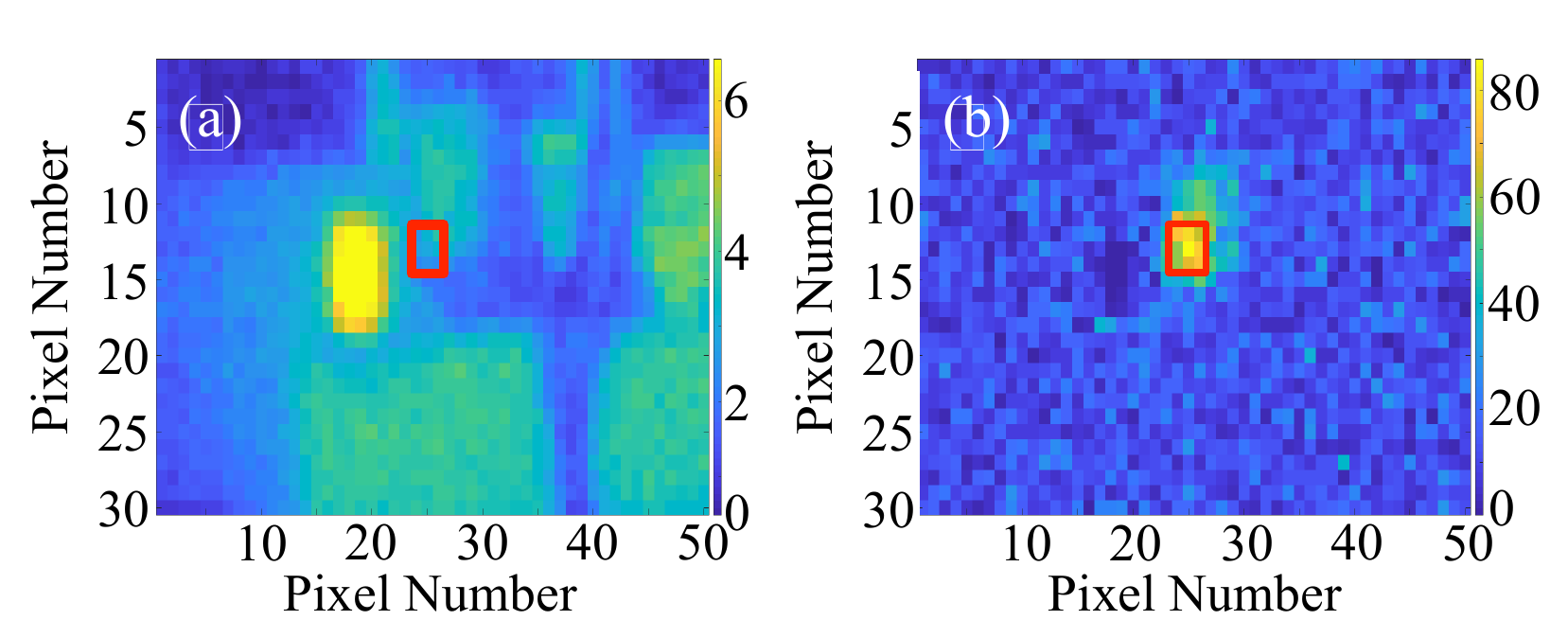}
\caption{Examples of raw and QLD-processed images acquired during day time. The red rectangle shows the position of the piece of cardboard that is illuminated by the modulated LED panel. (a) One full-scale raw frame (5~ms exposure time), showing reflection of sunlight from a polystyrene object located close to the LED panel (b) Corresponding processed image obtained from 10,400 raw frames acquired at 160 frames per second. The modulation frequency is 16~Hz.}
\label{Fig07}
\end{figure*}

The preceding results of Figs.\,\ref{Fig04},  \ref{Fig05}, and \ref{Fig06} were obtained at day-break, when the LED panel was much brighter than the ambient light. We now show that QLD can also be useful in distinguishing a modulated beacon or an object illuminated by a modulated source, from surrounding sources of light, especially during day time when many objects can reflect the sun light and blind the observer. Figure \ref{Fig07} illustrates this capability, using a similar setup as in Fig.\,\ref{Fig06}, but obtained during day time  fog. Here, only a cardboard piece is illuminated by the  modulated light provided by the LED panel. Close to it, but shielded from the modulated source, is a polystyrene block that strongly scatters sunlight towards the camera. Both objects, the cardboard and the polystyrene block, are 150~m from the camera, and are viewed through fog  during daylight. In Fig.\,\ref{Fig07}(a), the parasitic sunlight reflection is clearly visible, while the illuminated cardboard is impossible to distinguish, as confirmed by a CNR measured to be equal to $-0.3$ for this image. This image also exhibits vague shapes in the foreground due to intervening trees. After QLD processing (see Fig.\,\ref{Fig07}(b)), all these spurious shining objects disappear, and we are left with a clearly visible image of the piece of cardboard illuminated by the LED panel, with a CNR which is now  equal to 2.4. 

\section{Conclusion}
In conclusion, we have shown that computational QLD processing of images obtained using a modulated LED source is a powerful tool, compatible with real-time processing, which could be very useful for many applications. The fact that such imaging can be performed by illumiating with simple LEDs and processing on an ordinary computer shows that it can potentially be implemented at low cost, which further paves the way to a broad range of applications. In particular, this technique has been  proven to be efficient in  improving the visibility of  beacons under heavy fog conditions, particularly at night, a situation that is commonly encountered during plane landing and takeoff. Moreover, we have shown that it is also capable of imaging a reflecting or diffusive object which is illuminated by the modulated source of light.  Thus, in the context of aircraft navigation, unlike modern instrument landing systems that merely guide an aircraft using radio waves, the QLD technique can provide the pilot a visual image of the scene that lies ahead, and in particular a realistic representation of the runway beacons. In  motor, rail or maritime navigation, apart from showing the path by means of beacons, the technique may be used to reveal obstacles in the path, that are otherwise hidden by fog.  The technique is  particularly interesting if one wants to be able to steer the direction of emission of the modulated light, like in the case of a lighthouse, whose range could thus be extended in heavy fog conditions.  Finally, we have shown that source modulation and QLD also proves to be interesting in the presence of daylight, because it permits one to distinguish the beacon or object of interest  from any surrounding source of light that could dazzle the observer. 

Finally, it is worth mentioning that an important perspective of the present work, in the context of above mentioned application,  consists in assessing its validity in the case of moving targets.

\medskip
\noindent\textbf{Funding.} Indo-French Center for the Promotion of Advanced Research (CEFIPRA/IFCPAR 4406). Department of Science and Technology (DST), India. International Emerging Action project "HINDI-BIO" (CNRS), France.

\section*{Acknowledgments}
The authors are happy to thank Prof. Rupamanjari Ghosh for having made the experiments possible at Shiv Nadar University. FB acknowledges the hospitality of Raman Research Institute.

\end{document}